# Automating Vessel Segmentation in the Heart and Brain: A Trend to Develop Multi-Modality and Label-Efficient Deep Learning Techniques

Nazik Elsayed[1,2,6,7], Yousuf Babiker M. Osman[1,2,7], Cheng Li[1,7], Jiong Zhang[5], and Shanshan Wang[1,3,4*]

[1] Pual C. Lauterbur Research Centre for Biomedical Imaging, Shenzhen Institute of Advanced Technology, Chinese Academy of Sciences, Shenzhen 518055, China.
[2] University of Chinese Academy of Sciences, Beijing 100049, China.
[3] Peng Cheng Laboratory, Shenzhen 518055, China.
[4] Guangdong Provincial Key Laboratory of Artificial Intelligence in Medical Image Analysis and Application, Guangzhou 510080, China.
[5] Institute of Biomedical Engineering, Ningbo Institute of Materials Technology and Engineering, Chinese Academy of Sciences.
[6] Faculty of Mathematical and Computer Sciences, University of Gezira, Sudan.
[7] Nazik Elsayed, Yousuf Babiker M. Osman, and Cheng Li contributed equally to the manuscript.
[*] Authors to whom any correspondence should be addressed.

E-mail: sophiasswang@hotmail.com

## Abstract

Cardio-cerebrovascular diseases are the leading causes of mortality worldwide, whose accurate blood vessel segmentation is significant for both scientific research and clinical usage. However, segmenting cardio-cerebrovascular structures from medical images is very challenging due to the presence of thin or blurred vascular shapes, imbalanced distribution of vessel and non-vessel pixels, and interference from imaging artifacts. These difficulties make manual or semi-manual segmentation methods highly time-consuming, labor-intensive, and prone to errors with inter-observer variability, where different experts may produce different segmentations from a variety of modalities. Consequently, there is a growing interest in developing automated algorithms. This paper provides an up-to-date survey of deep learning techniques, for cardio-cerebrovascular segmentation. It analyzes the research landscape, surveys recent approaches, and discusses challenges such as the scarcity of accurately annotated data and variability. This paper also illustrates the urgent needs for developing multi-modality label-efficient deep learning techniques. To the best of our knowledge, this paper is the first comprehensive survey of deep learning approaches that effectively segment vessels in both the heart and brain. It aims to advance automated segmentation techniques for cardio-cerebrovascular diseases, benefiting researchers and healthcare professionals.

Keywords: Cardiovascular Diseases, Cerebrovascular Diseases, Deep Learning, Medical Imaging, Multi-Modality, Vessel Segmentation.

## Abbreviation

| Abbreviation | Full Term |
|---|---|
| AI | Artificial Intelligence |
| DL | Deep Learning |
| AD | Alzheimer's disease |
| DSA | Digital Subtraction Angiography |
| MRA | Magnetic Resonance Angiography |
| CTA | Computed Tomography Angiography |
| ML | Machine Learning |
| CNN | Convolutional Neural Networks |
| CA | Conventional angiography |
| DRA | Digital Radiographic Angiography |
| DUS | Duplex ultrasound |
| IVUS | Intravascular Ultrasound |
| CT | Computed Tomography |
| MRI | Magnetic Resonance Imaging |
| CE-MRA | Contrast Agent-Enhanced Magnetic Resonance Angiographic |
| NCE-MRA | Non-contrast Magnetic Resonance Angiographic |
| TOF-MRA | Time-Of-Flight Magnetic Resonance Angiography |
| PC-MRA | Phase Contrast Magnetic Resonance Angiography |

## I. INTRODUCTION

Cardio-cerebrovascular diseases, are the leading cause of death worldwide, responsible for over 17.9 million deaths in 2019, according to the World Health Organization [1], [2]. In addition, growing evidence indicates a strong relationship between dementia, particularly Alzheimer's disease (AD), and cardio-cerebrovascular illnesses [3]. In China, cardiovascular disease was the cause of 46.74% of deaths in urban regions and 44.26% of deaths in rural areas in 2019, according to the 2021 report on Cardiovascular Health and Diseases [4]. Furthermore, the report states that the number of individuals experiencing hypertension, stroke, and coronary heart disease in 2019 was 245.00 million, 13.00 million, and 11.39 million, respectively [4], [5].

The cardio-cerebrovascular system, which is a complicated network of arteries, and veins, carries nutrients and oxygen to the heart and brain. The small inter-individual variations in blood vessels are still not completely understood [6]. Medical imaging techniques such as digital subtraction angiography (DSA), magnetic resonance angiography (MRA), and computed tomography angiography (CTA) are effective tools for diagnosis, surveillance, and treatment planning for cardio-cerebrovascular diseases [7], [8]. Segmenting vessels in medical images is essential in diagnosing cardio-cerebrovascular diseases, as it helps specialists to isolate the arteries and veins from the surrounding tissue, allowing for better visualization and quantitative analysis. Nonetheless, vascular segmentation is still mainly achieved manually or semi-manually in the clinic. Both approaches are very time-consuming and labor-intensive and may produce errors due to the complicated topology and geometry of vessels , anatomical differences between patients, deficiencies caused by imaging techniques, and problem of data imbalance [3]. As a result, studies have focused on creating more efficient and precise automated vessel segmentation methods [6]. For this purpose, numerous classic vascular segmentation techniques[9][10] have been developed over the past years. These methods take advantage of various vascular image characteristics, including vessel intensity distributions, geometric models, and vessel extraction approaches. However, these techniques mostly rely on handcrafted features, manual parameters selection, and many processing steps [11].

Over the past few years, there has been a significant increase in the use of artificial intelligence (AI) methods, such as machine learning (ML) and deep learning (DL), for the early identification and diagnosis of cardio-cerebrovascular diseases. Specifically, DL approaches, such as convolutional neural networks (CNN), have exhibited significant advancements in accurately segmenting blood vessels. This improvement can be attributed to their enhanced utilization of contextual information and their ability to extract high-level features [6]. Instead of relying on manual feature engineering as is the case with conventional methods, DL can automatically extract blood vessel features and iteratively adapt them through training, leading to improved segmentation results. DL-based algorithms for cardio-cerebrovascular segmentation have gradually gained widespread adoption and outperformed conventional techniques [11], [12].

Several review papers have examined the segmentation of blood vessels. For example, Lesage et al. [13] reviewed traditional methods based on three aspects: models, features, and extraction schemes, with a specific emphasis on 3D contrast-enhanced imaging techniques such as MRA and CTA. However, their review did not cover the latest advancements in deep learning. In a different study, Moccia et al. [14] provided a comprehensive review of blood vessel segmentation algorithms in medical imaging. This review encompassed the segmentation of blood vessels across various anatomical regions and imaging modalities. However, the focus was primarily on discussing traditional methods, and fully supervised deep learning approaches. Similarly, Chen et al.[15] conducted a survey focusing on deep learning approaches like U-Net and other CNN-based models for cerebrovascular segmentation. However, none of these reviews have presented a systematic overview specifically focused on DL models for cardio-cerebrovascular segmentation. Given the significant advancements in this field, updated reviews are necessary to analyze and summarize the current state of the art.

The aim of this article is to provide a thorough examination of the existing DL techniques utilized in the segmentation of blood vessels within the heart and brain (Fig. 1). Specifically, the challenges encountered by these methods, such as handling complex shapes and variations, in addition to issues with limited annotated data, are discussed. In order to address these challenges, this work proposes the development of multi-modality label-efficient deep learning techniques as a promising future direction.

The structure of this paper is as follows: Section II provides a brief overview of the imaging modalities used for visualizing cardio-cerebrovascular structures, and the challenges of cardio-cerebrovascular segmentation. Section III provides a summary of recent DL-based approaches to cardio-cerebrovascular segmentation, respectively. Section IV summarizes available open-source datasets for cardio-cerebrovascular segmentation. In Section V, the urgent need for developing multi-modality label-efficient deep learning techniques is discussed. Finally, the last section concludes this paper.

## II. MULTIMODAL IMAGING FOR CARDIO-CEREBROVASCULAR VISUALIZATION

Despite the availability of numerous imaging modalities, angiography is the most effective technique for imaging blood vessels. In order to diagnose and assess the prognosis of cardio-cerebrovascular diseases, angiograms provide important information regarding blood flow and vascular system parameters. Each modality has distinctive features, which are described in detail below.

### *A. Conventional Angiography*

Conventional angiography (CA) is an invasive procedure utilizing radiographic imaging and contrast material injected directly into targeted blood vessels. Digital subtraction angiography (DSA) captures pre- and post-contrast image sequences, subtracting them to isolate blood vessel visualization, offering high spatial resolution but carrying drawbacks such as ionizing radiation exposure, procedure complexity, cost, and potential allergic reactions [13].

On the other hand, digital radiographic angiography (DRA) focuses on blood vessel visualization without requiring image subtraction like DSA. It is effective for digital image generation, processing, and archiving. Despite diagnostic advantages, DRA faces challenges including ionizing radiation use, potentially lower image detail, and cost concerns [16].

### *B. Ultrasound Duplex or Doppler ultrasound (DUS)*

DUS combines pulsed-wave Doppler and B-mode ultrasound, generating detailed images of body structures like organs, tissues, and blood vessels. It measures blood flow velocity using the Doppler effect but might be operator-dependent, varying in quality due to the operator's skill. While non-invasive and cost-effective, it might struggle with imaging small vessels [17], [18].

Intravascular ultrasound (IVUS) involves using miniaturized ultrasound probes inserted through catheters into the heart and blood vessels, providing real-time cross-sectional images. This technique offers detailed insights into arterial and cardiac structures, aiding in the detection of blockages and abnormal vessel conditions, and allowing for comprehensive analysis of cardiovascular anatomy and blood flow characteristics in a minimally invasive manner [19], [20].

### *C. Computed Tomography Angiography (CTA)*

Computed Tomography (CT) is a non-invasive medical imaging technology that utilizes X-rays to produce images of different organs. Similar to other modalities, angiography techniques using CT enable the capture of images of blood vessels. During CTA, thin section arterial phase CT images are taken after the patient receives an intravenous injection (IV) of iodinated contrast [15], [21], [22].

CTA has various advantages compared to DSA. For instance, CTA is typically less expensive, making it a more affordable option for patients and healthcare providers. Furthermore, CTA often requires shorter scan periods than DSA, resulting in a more efficient imaging process. One significant advantage of CTA is its ability to visualize not only the blood vessels but also the vessel walls and surrounding soft tissues. However, the main disadvantages of CTA include the use of potentially renotoxic contrast material and the patient's exposure to ionizing radiation [21], [22].

### *D. Magnetic Resonance Angiography (MRA)*

MRA is a widely used non-invasive imaging technique that utilizes magnetic fields and radio waves to visualize and evaluate blood arteries (Fig. 3). It generates high-resolution images able to help diagnose and assess various vascular disorders, such as cardio-cerebrovascular diseases [23]. MRA offers several benefits. Firstly, it is a completely noninvasive procedure. Additionally, MRA utilizes a contrast agent that is no nephrotoxic and has a lower risk of causing allergic reactions compared to other agents. Moreover, MRA avoids the use of ionizing radiation and eliminates potential complications associated with it. Finally, MRA offers the advantage of integrated display, allowing for the simultaneous visualization of both blood vessels and soft tissues [21].

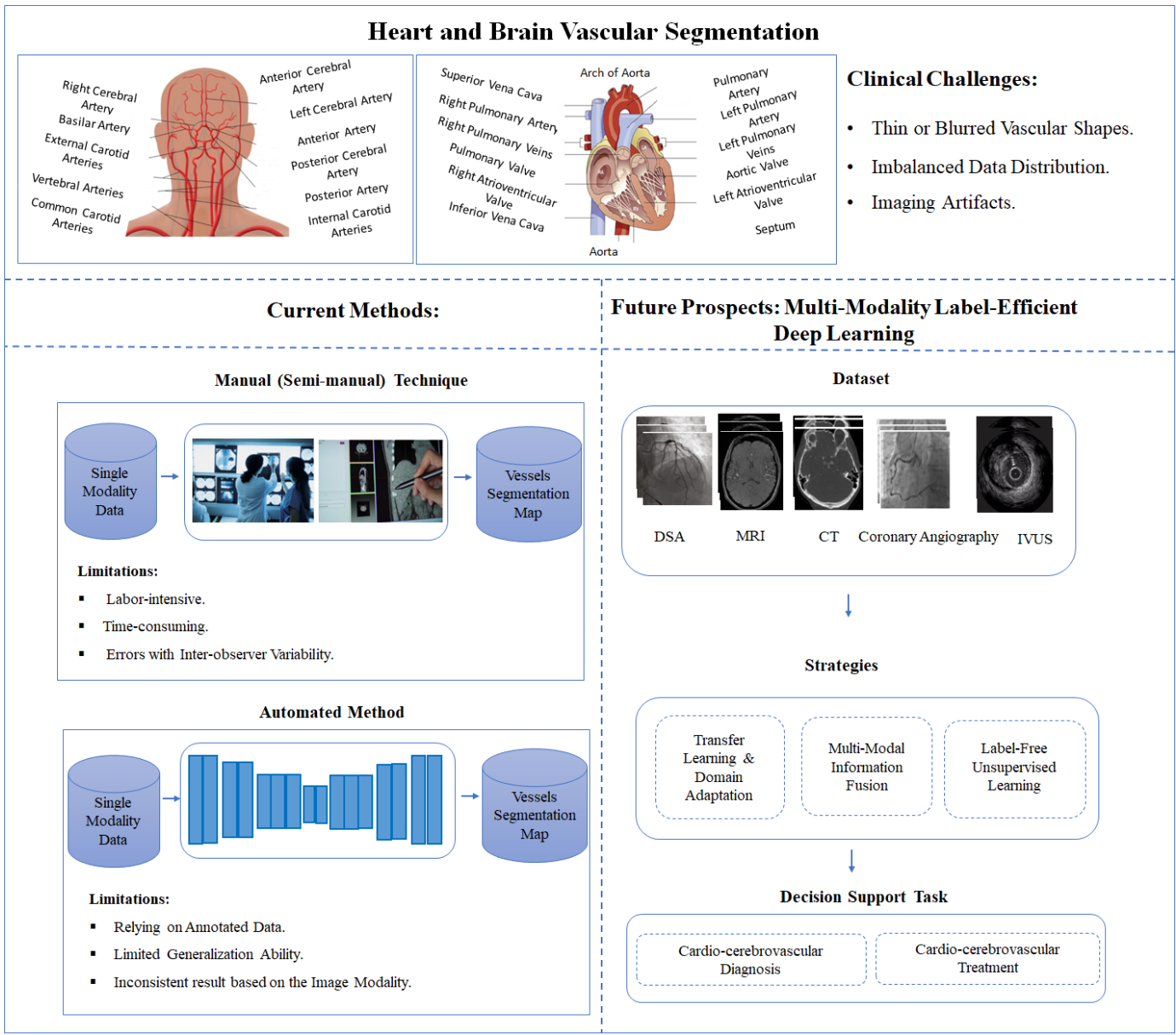

Fig. 1. Graphical overview of challenges and advances in the heart and brain vascular segmentation.

MRA techniques can be classified into two primary categories: contrast agent-enhanced MRA (CE-MRA) and non-contrast MRA (NCE-MRA) approaches. The NCE-MRA methods encompass time-of-flight MRA (TOF-MRA) and phase contrast MRA (PC-MRA) (Fig. 2) [17], [23].

### E. Others Modalities

Coronary angiography, an invasive method used to examine blood flow in the heart's arteries, provides accurate diagnosis and potential concurrent treatment, minimizing complications. Despite its effectiveness, it faces limitations such as suboptimal visualization due to technical constraints, variability in interpretation, the inability to observe vessel linings, and challenges in clot assessment [24].

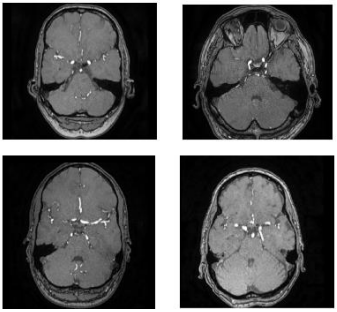

Fig. 2. Represents an example of TOF-MRI image from Aneurysm Detection and Segmentation (ADAM) dataset, http://adam.isi.uu.nl/.

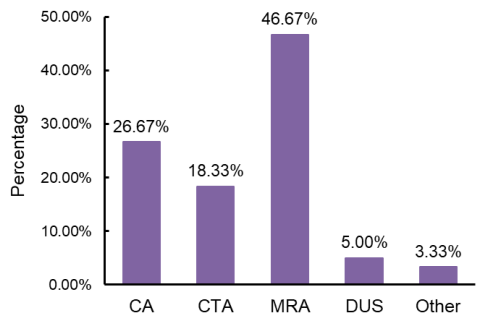

Fig. 3. Represents the distribution of modalities used in the reviewed studies.

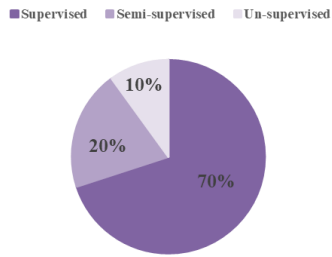

Fig. 4. Percentages of reviewed studies utilizing Supervised, Semi-supervised, and Un-supervised methods.

### F. Challenges in Segmenting Cardio-Cerebrovascular Structures

Segmenting cardio-cerebrovascular structures from medical imaging data can be a challenging task due to several factors. Some of the major challenges associated with this segmentation task are listed as follows.

#### 1) Complex Anatomy and Shape Variability

The cardio-cerebrovascular system is a complex network that consists of the heart, blood vessels, and the brain. These structures have various shapes, sizes, and patterns of connectivity. Among them, blood vessels are elongated tube-like structures that have a distinct directionality and anisotropy. It is crucial to accurately identify and separate blood vessels from medical images in order to detect and diagnose diseases related to the cardio-cerebrovascular system [3], [11], [25].

However, accurately delineating blood vessels within intricate anatomical scenes can be difficult due to the variations in vessel size, shape, and connectivity. Furthermore, the appearance and geometry of blood vessels can be affected by stents, calcifications, aneurysms, and stenosis, which can cause disruptions in the continuity of vascular networks. In medical imaging, differentiating blood vessels from surrounding anatomical structures, such as organs or bones, can be challenging. Obtaining adequate contrast between blood vessels and adjacent tissues becomes particularly difficult in complex anatomical scenes where there may be overlapping or adjacent tissues with similar radiographic properties [26], [27].

#### 2) Image Quality and Noise

Medical images that are utilized for cardio-cerebrovascular segmentation frequently exhibit noise, motion artifacts, low contrast, and imaging artifacts specific to particular imaging modalities [13]. These elements can pose challenges in obtaining high-quality images of blood vessels. Moreover, complicated anatomical scenes can cause extra sources of image degradation, making it challenging to distinguish between background artifacts and vessel structures. In addition, small vessels often exhibit weak intensities and low contrasts, making them susceptible to the influence of noise and the surrounding background [8], [5].

#### 3) Data Imbalance Issues

Data imbalance is a prevalent challenge in medical image segmentation tasks, including the segmentation of blood vessels. It occurs when the number of samples belonging to different classes is significantly unequal, resulting in an uneven distribution of data. In the case of blood vessel segmentation, vessels are thin structures that occupy only a sparse subset of pixels. This results in a severe imbalance between the number of pixels representing vessels and those representing non-vessel areas [28].

#### 4) Scarcity of accurately annotated data

A significant challenge to training DL models for cardio-cerebrovascular segmentation is the lack of diverse, high-quality, and accurately annotated training datasets. Privacy concerns related to patient data restrict the accessibility and sharing of medical image datasets. Moreover, the collection and annotation of data is a resource-intensive, and time-consuming process, adding to the difficulty of acquiring diverse training datasets. Annotating medical images for cardio-cerebrovascular segmentation requires expertise and manual effort, which further amplifies the challenge of obtaining annotated datasets on a large scale [11], [29], [30].

## III. DEEP LEARNING VESSEL SEGMENTATION APPROACHES FOR THE BRAIN AND HEART

DL approaches have demonstrated substantial promise in automating the cardio-cerebrovascular structure segmentation process and boosting the segmentation reliability. DL approaches could be broadly classified as Label-intensive fully-supervised and Label-efficient DL approaches.

### A. Label-Intensive Fully-Supervised Approaches

Fully-supervised DL methods have achieved high accuracy for cardio-cerebrovascular segmentation tasks. These methods were trained end-to-end in a supervised manner using manually segmented images. In this section, we categorize the complete set of supervised methods for cardio-cerebrovascular segmentation into seven types based on their models or design approaches (Table I).

#### 1) U-Net Model

Recently, convolutional neural networks (CNNs) have been widely used for medical segmentation tasks, including cardio-cerebrovascular segmentation. One notable architecture, U-Net, utilizes an encoder-decoder structure with skip connections. This allows for the extraction and refinement of low-resolution features, leading to impressive results in various segmentation tasks. Numerous variations of U-Net have been proposed specifically for cardio-cerebrovascular segmentation [31].

For example, Mu et al. [32] proposed a deep encoder-decoder architecture with preprocessing and post-processing for intracranial aneurysm (IA) segmentation. The proposed work constructed a multi-scale supervision-based attention residual U-Net (ARU-Net), incorporating a depth-aware attention gate module to improve the learning of accurate IA structures and small vessels, and utilized a 3D conditional random field (3D CRF) to eliminate unwanted adhesions/connections between adjacent arteries and IA in the segmentation results. Furthermore, Meng et al. [33] presented a CNN-based segmentation framework: called the multiscale dense CNN, which aims to automatically segment cerebral vessels in DSA images. The framework incorporates redesigned skip connections between the encoder and decoder stages to improve the performance of the segmentation task diameters. Alidoost et al. [34] suggested to utilize a CNN to automatically segment brain blood vessels in order to determine whether the results satisfy radiologists' criteria for disease diagnosis. To extract brain vasculatures from volumetric MRA images, the study utilized a deeply supervised attention-gated 3D U-Net trained with the focal Tversky loss function. Moreover, Li et al. [35] suggested a segmentation network to separate cerebral vessels from MRA images. In their network, a global vascular context module captures the long-distance reliance between vascular structures, while the hybrid loss function ensures excellent connectivity by constraining the topology during the segmentation process. Fu et al. [36] developed a computational tool for reconstruction of CTA images. The tool can be utilized to automatically segment and extract entire vessels from head and neck scans.

A generic neural network called ER-Net to segment vessel-like structures in various 3D medical imaging modalities was proposed by Xia et al. [37]. To enhance the identification and preservation of spatial edge information, a reverse edge attention module and an edge-reinforced optimization loss were developed. Also, Song et al. [38] proposed a two-step automatic method for coronary artery segmentation in coronary CTA images using a deep CNN. The first step involved the use of a 2D DenseNet classification network to identify non-coronary-artery slices, while in the second step, accurate coronary artery segmentation was achieved using a 3D-UNet network with feature fusion and rectification.

In order to overcome the structural simplicity of U-Net, Jun et al. [39] proposed a framework called T-Net, which is a novel nested encoder-decoder architecture that has been introduced for main vessel segmentation in coronary angiography. It offers a more accurate and efficient approach for feature connection and pooling/up-sampling arrangement. By leveraging this architecture, T-Net provides enhanced precision and effectiveness in the task of segmenting main vessels in coronary angiography. Sieren et al. [40] proposed a DL-based algorithm for segmenting and quantifying the physiological and diseased aorta in CTA. The framework was able to achieve high accuracy, even in cases where the aorta was diseased or the vascular architecture was distorted.

#### 2) Other Encoder-Decoder Models

In addition to the U-Net-based model, several other models have been developed and reported for cardio-cerebrovascular segmentation. Utilizing CTA volumes in regions around the skull, Nazir et al. [41] suggested an optimally fused fully end-to-end network (OFF-eNET) for automatically segmenting dense volumetric 3D cerebral vascular structures.

Additionally, Fu et al. [42] proposed the use of deep convolutional neural networks for automatic detection of blood vessels in cerebral angiograms. This approach can improve efficiency and allow for the development of automatic diagnosis systems. Phellan et al. [1] developed a deep CNN aimed at automatically segmenting the vessels of the brain in TOF MRA images from healthy individuals. The network was trained using manually annotated bi-dimensional image patches sourced from axial, coronal, and sagittal directions. Furthermore, Dumais et al. [43] developed a DL framework for automatically segmenting, labeling, and quantifying circle of Willis (CW) arteries on MRA images. The CNN was trained on a dataset of MRA images with CW arteries. The model was able to achieve a high accuracy in segmenting, labeling, and quantifying the CW arteries, even in cases where the images were of poor quality. Also, Dou et al. [44] presented a novel approach for automatically detecting cerebral microbleeds (CMBs) from MR images using 3D CNNs. Kandil et al. [45] proposed a computer-aided diagnosis system for the early detection of hypertension by analyzing MRA data of human brains. The proposed system can accurately model and discriminate between normal and hypertensive subjects, which can help clinicians prepare appropriate treatment plans to mitigate adverse events.

For coronary artery centerline extraction at any given points in coronary CTA, Wolterink et al. [46] presented a deep learning-based technique based on a local image patch. In a limited amount of training images, the approach can be properly trained through manually labeled reference centerlines. Also, Tetteh et al. [29] proposed a DeepVesselNet, a DL architecture for vessel segmentation, centerline prediction, and bifurcation detection in 3D angiographic volumes. The architecture addressed challenges such as low execution speed, high memory requirements, high-class imbalance, and unavailability of accurately annotated 3D training data.

In addition, Zhu et al. [47] developed a deep learning framework for the accurate extraction of vascular structures from vascular images. The framework employed computer vision and the PSPNet network, and it was shown to be more accurate than traditional algorithms, and it can reduce manual interaction in diagnosis. Moreover, Samuel et al. [48] introduced a method for segmenting blood vessels from retinal fundus and coronary angiogram images, leveraging a single VSSC Net with improved accuracy.

### 3) General Adversarial Network (GANS)

The application of GANs in medical image segmentation, including blood vessel segmentation, has gained significant attention due to their ability to generate and identify images. Researchers have actively explored GANs to improve segmentation accuracy. For example, Chen et al. [27] developed A-SegAN, a 3D adversarial network model aimed at segmenting cerebral vessels in TOF-MRA 3D images. This model consists of a segmentation network, which generates segmentation maps, and a critic network, which is responsible for discriminating between the predictions and the ground truth. The proposed model incorporates a visual attention mechanism to enhance its segmentation performance.

Furthermore, Subramaniam et al. [49] introduced a 3D GANs model that effectively generated TOF-MRA patches while also providing brain blood vessel segmentation labels. They utilized four variations of 3D Wasserstein GAN (WGAN) to achieve this, which encompassed: 1) gradient penalty (GP), 2) gradient penalty with spectral normalization (SN), 3) spectral normalization with mixed precision (SNMP), and 4) spectral normalization and mixed precision with double filters per layer (c-SN-MP).

Similarly, Chen et al. [50] introduced a GAN-based model which employs a hybrid loss function, integrating dice loss with least square loss, for model training. This model, designed for effectively segmentation of cerebrovascular structures in TOF-MRA data, utilized the SEVnet structure as the generator.

Additionally, Guo et al. [51] developed a deep learning framework for cerebrovascular segmentation on MRA images. The framework, called FiboNet-VANGAN, was an end-to-end adversarial model that can address the challenges of large variations in vascular anatomies and voxel intensities, severe class imbalance between foreground and background voxels, and image noise with different magnitudes. The framework was able to achieve high segmentation accuracy on a dataset of MRA images.

Utilizing sparse labeling, Kossen et al. [52] developed a generative adversarial networks to create synthetic neuroimaging data and corresponding labels for arterial brain vessel segmentation. The synthesized image-label pairs were shown to retain generalizable information and can be utilized in a transfer learning approach with independent data, paving the way to overcome the challenges of scarce data and anonymization in medical imaging.

### 4) Fusion based Models

Fusion-based models are used in various applications, including cardio-cerebrovascular segmentation. They integrate information from multiple sources to improve segmentation accuracy. Fusion can occur at different levels—feature, decision, or intermediate—by combining features, segmentation results, or information at intermediate stages of segmentation.

For instance, A multiple-U-net (M-U-net) architecture, which trains three U-nets in three directions, was designed as the basis of the automatic cerebrovascular segmentation algorithm presented by Guo et al. [53]. A voting mechanism was employed for feature fusion after receiving the heat maps generated by the three single U-net models. In order to obtain the final results of cerebrovascular segmentation, connected component analysis was employed as the post-processing step.

Similarly, Zhao et al. [25] developed a deep neural network for vessel segmentation to improve the accuracy of vessel reconstruction for vascular disease diagnosis. The network consists of two sub-networks, with the second sub-network having two components, a traditional CNN-based U-Net and a graph U-Net, which perform cross-network multi-scale feature fusion to support high-quality vessel segmentation.

In addition, Shi et al. [3] introduced a method called the affinity feature strengthening network (AFN), which combines geometric modeling and enhancement of pixel-wise segmentation features through a contrast-insensitive, multiscale affinity technique.

### 5) Transformer based Models

Transformer-based models have attracted considerable interest in various domains such as natural language processing and computer vision. Recently, researchers have started investigating their application in blood vessel segmentation tasks. Transformers are distinguished by their unique architecture that incorporates self-attention mechanisms, which improve the correlation between different parts of the input data [54], [8].

TABLE I
PART OF THE FULLY-SUPERVISED AND LABEL- EFFICIENT DEEP LEARNING STUDIES ON CARDIO-CEREBROVASCULAR SEGMENTATION.

| References | DL Method | Vessel Target | Modality | Data Availability | Code Availability |
| --- | --- | --- | --- | --- | --- |
| Mu et al 2023 | Fully-Supervised | Cerebrovascular | 3DRA (23 3DRA images (65% 3DRA images were used as training)). Images were manually annotated by experts. | Public | No |
| Shi et al 2023 | Fully-Supervised | Cardiovascular | XRA (126 subjects, 84 for training, and 42 for testing). Labels manually annotated by experts. | Public | Yes |
| Alidoost et al 2023 | Fully-Supervised | Cerebrovascular | MRA (IXI dataset (68 volumetric MRA images of healthy subjects were used. | Public | No |
| Zhao et al 2023 | Fully-Supervised | Cardiovascular | CTA (Automated Segmentation of Coronary Arteries Dataset (ASOCA) dataset, Aorta and Coronary Artery Dataset (ACA) (1000 CCTA images (800 training set, 100 validation set and 100 test set))). Head and Neck Artery Dataset (HNA) 800 CTA images (640 training set, 80 validation set and 80 test set). Labels were annotated by experts. | Public | No |
| Pu et al 2023 | Semi-supervised | Cardiovascular | DSA (It contains 300 sets of coronary X-ray angiographic sequences from 50 patients). | Private | Yes |
| Chen et al 2023 | Semi-supervised | Cerebrovascular | TOF-MRA (MIDAS dataset (95 cases of manual cerebrovascular results from experts as ground truth were used (10 sets of data was used as testing set))). | Public | Yes |
| Gu et al 2023 | Semi-supervised | Cardiovascular | XRA (191 XAs of 30 patients). | Private | Yes |
| Dang et al 2022 | Semi-supervised | Cerebrovascular | TOF-MRA (Three different types of data were used: synthetic (136 images), Time-of-Flight (TOF) angiography (100 images) & Susceptibility-Weighted Images (SWI) (33 images). | Public | Yes |
| Zhang et al 2023 | Un-supervised | Cardiovascular | XRA (260 clinical XCA sequences were collected. The dataset is divided into training, validation and testing at a ratio of 136:60:64). Each sequence was annotated by two clinicians to obtain the three key-frame locations. | Public | Yes |
| Zhou et al 2023 | Un-supervised | Cardiovascular | US (The SPARC dataset: 510 US images of carotid plaques were collected from 144 patients. The Zhongnan dataset: 638 images were collected from 497 patients). | Private | No |
| Huang et al 2022 | Un-supervised | Cerebrovascular | CTA (24 patients (16 patients had atherosclerosis with calcified plaques or mild or moderate stenosis in various arteries & 4 patients had thinner vessels in the vertebral artery or the anterior arteries. Additionally, 1 patient had intracranial hemorrhage, and 1 patient had a craniopharyngioma. No obvious lesion was found in three patients). | Private | No |

In order to develop rich hierarchical representations of curvilinear structures, Mou et al. [54] introduced a new curvilinear structure segmentation network (CS2-Net), which contains a self-attention mechanism in the encoder and decoder.

Hao et al. [8] developed a DL framework that can automatically segment contrast-filled vessels from X-ray coronary angiography (XCA) image sequences. The framework used a channel attention mechanism to focus on the most important features in each frame, which helps to improve the accuracy of the segmentation. Also, Zhang et al. [55] developed a DL framework for direct quantification of coronary artery stenosis on X-ray angiography images. The framework utilized a hierarchical attention mechanism to learn from multiple views of the coronary arteries. This helps to improve the accuracy of the quantification, especially in cases where the stenosis is overlapping.

Moreover, Ni et al. [56] presented a network for intracranial vessel segmentation in CTA called GCA-Net (Global Channel Attention Network). GCA-Net used a four-branch shallow feature extractor to efficiently capture global context information, which helps to improve the segmentation accuracy. GCA-Net also focused on preserving more feature details, which helps to improve the robustness of the segmentation to noise and other artifacts.

**6) Graph Convolutional Networks**

Graph Convolutional Networks (GCNs) have demonstrated potential in diverse graph-related tasks, such as blood vessel segmentation. GCNs utilize graph theory to represent images as graphs, with nodes representing pixels or image regions and edges capturing their connections or interactions. This approach allows GCNs to capture the relationships and dependencies between different elements of the image, leading to effective segmentation results [46], [57].

For the segmentation of coronary arteries in cardiac CTA, Wolterink et al. [46] suggested an approach based on graph convolutional networks. Both local image features and features of adjacent vertices in the mesh graph were employed to make predictions about individual vertex positions.

Additionally, Shin et al. [57] presented a novel DL system for vessel segmentation that combines a graph neural network with a CNN. This approach can jointly exploit both local and global vessel information to produce more accurate segmentation results.

### 7) Other Models

Cui et al. [58] introduced a method for coronary artery lumen segmentation that utilizes supervised machine. The approach focuses on achieving high accuracy while minimizing user interaction. It involves developing a fully discriminative lumen segmentation technique that simultaneously learns a classifier relied upon by weak learners and the features of the classifier learning.

## B. Label-Efficient Deep Learning Approaches

Label-efficient DL approaches such as semi-supervised and un-supervised designed to optimize learning models by reducing their reliance on annotated data during the training phase. These approaches aim to achieve high performance even with limited or no labeled samples, thus minimizing the need for extensive manual annotation, which can be resource-intensive and time-consuming (Table I) [59].

### 1) Consistency-based Semi-Supervised Approaches

Consistency-based methods ensure predictions remain consistent under perturbations. For example, Chatterjee et al. [60] developed a deep learning method that employed the U-Net multi-scale supervision architecture for the automatic segmentation of small vessels in 7 Tesla 3D TOF-MRA data. Training the model on a limited, semi-automatically segmented dataset, which was imperfect, enabled it to improve its generalization performance by being made equivariant to elastic deformations through a self-supervised deformation-aware learning approach.

Furthermore, Chen et al. [61] proposed a semi-supervised generative consistency (GCS) model for cerebrovascular segmentation from TOF-MRA. The GCS model constrains the segmentation model using the generation results and improves the feature mining ability by calculating the consistency of the perturbed data. Xie et al. [62] proposed a semi-supervised cerebrovascular segmentation method for TOF-MRA images that achieves high performance by incorporating unlabeled data. Their method, which uses a region-connectivity-based mean teacher model (RC-MT), outperformed competing semi-supervised methods.

Similarly, Dang et al. [63] introduced a deep learning-based framework for segmenting 3D brain vessel images. The framework reduces the annotation time by approximately 77% by avoiding pixel-wise annotations instead only requiring weak patch-level labels to train the segmentation network. Pu et al. [64] proposed a semi-supervised learning method for coronary vessel segmentation that leverages the strengths of both CNNs and transformer models. The method achieved state-of-the-art performance with a small amount of labeled data by using pyramid-consistency learning and confidence learning to regularize the model and improve its generalization ability. Also, Yin et al. [65] developed a method for estimating optical flow in coronary angiography sequences using semi-supervised learning. The approach leveraged a combination of original medical images, region of interest segmentation, and pre-trained models from other optical flow datasets to train a new model suitable for medical imaging applications.

Moreover, Zhang et al. [66] presented a weakly supervised training framework that diverged from the conventional reliance on accurate labels obtained through manual annotation. Instead, the framework leveraged noisy pseudo labels generated via automatic vessel enhancement. To mitigate the impact of label noise and systematic biases in pseudo labels, a self-paced learning scheme was incorporated, ensuring the training process remains robust and effective.

Gu et al. [30] proposed a method called CS-CADA, which leveraged existing annotated datasets with similar anatomical structures to improve medical image segmentation performance. The method required limited annotations in the target domain and employs domain-specific batch normalization to learn domain-invariant features individually.

### 2) Minimization-based Semi-Supervised Approaches

Entropy minimization approaches incorporate labeled and unlabeled data by minimizing objective functions.

For instance, Zhang et al. [12] developed a strategy to automate cerebrovascular segmentation from TOF-MRA data. The proposed method integrates model- and data-driven methods and achieves better completeness and sensibility for slender vascularity.

In addition, Vepa et al. [67] suggested a learning framework for cerebral DSA vascular segmentation employing weak supervision and affordable human-in-the-loop techniques. In addition to a consideration of trade-offs between annotation cost and model performance when employing weak supervision strategies, the suggested methodology yielded state-of-the-art results for cerebral DSA vascular segmentation that even surpassed the quality of human annotators.

### 3) Label-Free Unsupervised Approaches

Traditionally, DL models have relied on large labeled datasets to train. However, these datasets can be time-consuming and expensive to acquire. Unsupervised learning paradigms, such as self-supervised learning, have emerged as a way to reduce the reliance on human annotations [5], [68]. Self-supervised learning methods can learn spatial representations by predicting the relative positions between two image patches. This task is a form of self-supervision, because it does not require any labeled data. This makes it a scalable and efficient way to learn representations from large unannotated datasets.

Fan et al. [68] developed a DNN-HMRF framework for unsupervised cerebrovascular segmentation of TOF-MRA images. The framework leverages the strengths of both DNN and hidden Markov random field (HMRF) to train the model with a limited number of annotations, resulting in improved segmentation performance. Ma et al. [5] proposed a solution to the challenge of annotating vessel segmentation maps in medical images, using a self-supervised approach based on adversarial learning by training an attention-guided generator and a segmentation generator to simultaneously synthesize fake vessels and segment vessels out of coronary angiograms, which forces them to learn accurate representations of vessels. Zhang et al. [69] introduced an approach for key frame localization in X-ray coronary angiography by integrating a convolutional long short-term memory (CLSTM) network into a multiscale attention Transformer. This integration enables the learning of segment and sequence-level dependencies within consecutive-frame-based deep features and image-to-patch contrastive learning. Huang et al. [59] suggested a self-supervised dual-task DL approach to automatically segment all vessels and forecast unenhanced CT images from single-energy head-neck CTA, leveraging the correlation between the two tasks to enhance performance without manual annotation. Zhou et al. [70] presented an image reconstruction-based self-supervised learning algorithm for carotid plaque segmentation, which consist of downstream segmentation task and pre-trained task. By reconstructing plaque images from randomly partitioned and disordered images, the pre-trained task learns region-wise representations with local consistency. The downstream task then transfers the initial parameters from the pre-trained model to the segmentation network.

TABLE II
SUMMARY OF THE OPEN-SOURCE DATASETS

| Datasets Name | Type & Size of Dataset | Image Acquiring Description & Resolution | Link |
|---|---|---|---|
| MIDAS Dataset | -TOF-MRA(95) -Size (448 × 448 × 128 voxels) & 14 volumes (352 × 448 × 176 voxels). | -SIEMENS ALLEGRA 3.0T MRI (TR=35.0, TE=3.56, and flip angle=22). -Resolution (0.5 × 0.5 × 0.8 mm³). | https://imed.nimte.ac.cn/MMM-21.html |
| Cerebrovascular Segmentation | -TOF-MRA(45) – -Size (1024 × 1024 × 92). | -1.5T GE MRI. - Resolution (0.264 × 0.264× 0.8mm3). | http://xzbai.buaa.edu.cn/datasets.html |
| TubeTKDataset | -T1 & MRA (100 healthy subjects) T1size (176×256×176) image & MRA. - Size (448×448×128). | -Siemens Allegra head-only 3T MR Scanner. - MRA Resolution (0.5 × 0.5 × 0.8mm) & T1 Resolution (1mm isotropic). | https://public.kitware.com/Wiki/TubeTK/Data. |
| ASOCA Dataset | -CTCA (Training set size 40(20 with coronary disease & 20 without) & Testing set size 20(10 healthy and 10 patients with disease)) -Size (512×512×N). | -a GE LightSpeed 64 slice CT scanner. - Anisotropic resolution with a 0.3–0.4 mm in-plane resolution and a 0.625 mm out-of-plane resolution. | https://asoca.grand-challenge.org |
| IXI Dataset | - [T1, T2 ,MRA , PD-weighted images, & Diffusion-weighted images] (600 normal, healthy subjects) | -a Philips 3T system at Hammersmith Hospital, a Philips 1.5T system at Guy's Hospital, and a GE 1.5T system at the Institute of Psychiatry | http://brain-development.org/ixi-dataset/ |
| Database X-ray Coronary Angiograms (DCA1) | -X-ray (30). -Size (300 ×300 pixels). | Not-provided | http://personal.cimat.mx:8181/~ivan.cruz/DB_Angiograms.html. |
| X-ray angiography Coronary Artery Disease (XCAD) | -Coronary angiography(training set (1621 mask frames &1621 coronary angiograms), & Testing set (126 t coronary angiograms) - Size (512 × 512 pixel). | Not-provided | https://github.com/AISIGSJTU/SSVS |
| Aneurysm Detection And segmentation (ADAM) | -TOF-MRA(255(113 Training set & 142 Testing set )) | Not-provided | http://adam.isi.uu.nl/ |

## IV. OPEN-SOURCE MULTI-MODALITY DATASETS FOR VESSEL SEGMENTATION

Datasets are essential for the development of DL models for the diagnosis of cardio-cerebrovascular diseases. Researchers have been provided with several datasets for cardio-cerebrovascular diagnosis, including MRA and CTA. In this section, we present a comprehensive overview of the crucial datasets available for cardio-cerebrovascular diseases, aiding researchers in their diagnostic endeavors (Table II).

## V. THE URGENT NEED FOR DEVELOPING MULTI-MODALITY LABEL-EFFICIENT DEEP LEARNING TECHNIQUES

The segmentation of blood vessels is vital for comprehending and diagnosing cardio-cerebrovascular diseases. Although existing segmentation methods have made notable advancements, it is imperative to explore new research avenues to overcome the aforementioned limitations and enhance the accuracy and efficiency of segmentation. This section presents a proposed solution to tackle challenges related to complex vessel shapes and limited availability of data.

### A. Label-Efficient Transfer Learning and Domain Adaptation

While the limited availability of training samples and the absence of annotations pose challenges in blood vessel segmentation, there are numerous datasets with comprehensive annotations available in related medical imaging tasks or general image domains. To address this, transfer learning techniques can be employed to utilize pre-trained models from these datasets to aid in training a model for blood vessels segmentation [71], [29]. Moreover, domain adaptation (DA) methods can be explored to enhance the generalization capabilities of models trained on one distribution (source domain) to other distributions (target domain) encountered in clinical settings. DA has gained attention in the field as a solution for minimizing the distribution gap among different yet related domains [30].

By leveraging knowledge from related domains or adapting models to specific target domains, these approaches enhance the model's ability to generalize well, even in scenarios with limited labeled data or domain shifts. This ultimately leads to more accurate and robust segmentation for blood vessels. One issue however, is the domain gap between the source and target datasets. How effectively transfer learning and domain adaptation heavily relies on the assumption that the source and target domains share similar characteristics. However, in medical imaging, the domain shift can be significant, especially when dealing with different imaging modalities, acquisition protocols, or patient populations. Bridging this domain gap remains a challenge and requires more robust techniques for effective adaptation.

### B. Multi-Modal and Multi-Scale Information Fusion

Cardio-cerebrovascular structures exhibit variations due to factors like imaging protocols, patient characteristics, and pathologies in scale, ranging from large vessels to fine branches. Multi-scale information integration allows models to capture structural details at different levels. This enhances the segmentation accuracy, particularly in complex anatomical regions with varying vessel sizes. Multi-modal and multi-scale integration provides contextual information that can aid in accurate segmentation. By considering the relationships and spatial dependencies between different scales and modalities, segmentation models can better understand the context of vessels within the surrounding anatomy, leading to improved segmentation results [72].

However, multimodal images usually face irregular data structures like physical space offsets and in consistencies in sampling parameters. Effectively addressing the fusion problem arising from irregular spatial structures among multimodal images is a key area for further exploration.

### C. Opportunities for More Powerful Label-Free Unsupervised Learning Approaches

Accurate and comprehensive labels for heart and brain blood vessel datasets can be costly and challenging. Unsupervised learning approaches emerge as potential solutions to mitigate the reliance on labeled datasets[5]. However, these approaches often require higher computational efficiency and training costs. Exploring more effective image features to improve the efficiency of unsupervised learning, is indeed a worthwhile direction to pursue, aligning with clinical needs. By identifying and leveraging informative and discriminative features from unlabeled data, these approaches can potentially improve segmentation accuracy and reduce the burden of manual annotation. While there have been some studies[68], [69] that have utilized unsupervised methods for cardio-cerebrovascular segmentation, there is still a need for additional research and development in this area (Fig. 4). Additional exploration and refinement of these techniques are necessary to improve the efficiency and cost-effectiveness of segmentation in the context of cardio-cerebrovascular imaging.

## VI. Conclusion

Cardio-cerebrovascular diseases pose significant challenges to global healthcare systems, necessitating accurate vessel delineation in the heart and brain for diagnosis and treatment planning. DL techniques have revolutionized automated segmentation in cardiac and cerebral imaging, excelling at discerning intricate vessel structures in medical images. This review provides a comprehensive summary of DL advancements in segmenting heart and brain blood vessels, addressing challenges and limitations of current methods. Urgent development of multi-modality label-efficient DL techniques is proposed to overcome issues like data scarcity and inter-observer variability. Such advancements can enhance segmentation accuracy, efficiency, and clinical applicability, benefiting both research and patient care. Continued efforts in refining these techniques will drive progress in automated vessel segmentation, enabling seamless integration into clinical workflows.

## CONFLICTS OF INTEREST

The authors have no potential conflicts of interest.

## ACKNOWLEDGEMENTS

This research was partly supported by the National Natural Science Foundation of China (62222118, U22A2040), Guangdong Provincial Key Laboratory of Artificial Intelligence in Medical Image Analysis and Application (2022B1212010011), and Shenzhen Science and Technology Program (RCYX20210706092104034, JCYJ20220531100213029), Key Laboratory for Magnetic Resonance and Multimodality Imaging of Guangdong Province (2023B1212060052).